\documentclass[fleqn,10pt,twocolumn]{wlscirep}
\usepackage[utf8]{inputenc}
\usepackage[T1]{fontenc}
\usepackage[utf8]{inputenc}
\usepackage{amsmath}
\usepackage[utf8]{inputenc}
\usepackage{amsmath}
\usepackage{algorithm}
\usepackage[noend]{algpseudocode}
\usepackage{setspace}
\usepackage{booktabs}

\title{Two Layer Walk: A Community-Aware Graph Embedding}

\author[1,2,*]{He Yu}
\author[1,2]{Jing Liu}

\affil[1]{School of Artificial Intelligence, Xidian University, 2 South Taibai Road, Xi’an, Shaanxi 710071, China}
\affil[2]{Guangzhou Institute of Technology, Xidian University, Knowledge City, Guangzhou, Guangdong 510555, China}

\affil[*]{corresponding author. yuhe001@stu.xidian.edu.cn}

\begin{abstract}
Community structures play a pivotal role in understanding the mesoscopic organization of networks, bridging local and global patterns. While methods like DeepWalk and node2vec effectively capture node positional and local structural information through random walks, they fail to incorporate critical community information. Other approaches, such as modularized nonnegative matrix factorization and evolutionary algorithm-based methods, preserve community structures but suffer from high computational complexity, making them unsuitable for large-scale networks. To address these limitations, we propose Two Layer Walk (TLWalk), a novel graph embedding algorithm that explicitly incorporates hierarchical community structures. By balancing intra- and inter-community relationships through a community-aware random walk mechanism automatically without using any parameters, TLWalk achieves robust and scalable representation learning that can fully extract local and global topologies, which is proved theoretically by showing TLWalk can overcome locality bias in the walk. We also theoretically prove the relationship between TLWalk and matrix factorization. Extensive experiments on benchmark datasets demonstrate TLWalk’s superior performance, with significant accuracy gains—up to 3.2\%—over existing methods for the link prediction task. TLWalk’s ability to encode both dense local and sparse global structures ensures its adaptability across diverse network types, offering a powerful and efficient solution for network analysis.

\end{abstract}

\begin{document}

\flushbottom
\maketitle
\thispagestyle{empty}

%\noindent Please note: Abbreviations should be introduced at the first mention in the main text – no abbreviations lists. Suggested structure of main text (not enforced) is provided below.

\section*{Introduction}

Graphs are ubiquitous in representing complex systems across diverse domains, from social networks \cite{Watts1998} and biological systems \cite{Jeong2000} to transportation \cite{Poorzahedy2005} and recommendation systems \cite{Koren2009}. These structures, consisting of nodes and edges, capture relationships and interactions within the system. Analyzing these graphs is crucial for tasks like link prediction, node classification, and uncovering hidden structures. Graph embedding provides a powerful solution by mapping graphs into low-dimensional vector spaces while preserving their essential properties \cite{perozzi2014deepwalk,grover2016node2vec,tang2015line, dall2024embedding,Zhang2021ConsistencyOR}. These embeddings encode positional relationships, structural roles, and patterns, facilitating efficient downstream tasks. Positional information reflects how nodes are situated, while structural roles highlight patterns and topological features. A wide range of embedding methods has been developed, broadly categorized into shallow and deep learning-based approaches. Shallow methods, such as DeepWalk \cite{perozzi2014deepwalk}, node2vec \cite{grover2016node2vec}, and matrix factorization \cite{Qiu:2018}, efficiently capture local and global patterns through strategies like random walks or linear decomposition. Deep learning approaches, such as Graph Neural Networks (GNNs) \cite{kipf2017semi} and Graph Attention Networks (GATs) \cite{velivckovic2018graph}, address non-linear dependencies but face challenges like over-smoothing, high computational demands, and sensitivity to hyperparameters.

Communities, residing at the mesoscopic level, bridge microscopic node-level and macroscopic graph-level structures \cite{newman2004finding, girvan2002community,fortunato2022years,Teng2021Overlapping,Teng2021Synchronous}. By identifying densely connected groups of nodes, communities provide a unique lens for understanding intermediate network organization. Incorporating community information into embeddings enhances their capacity to encode local and global patterns, making them vital for nuanced network analysis. Despite these advancements, effectively incorporating community structure into graph embeddings remains a critical challenge. Early community-preserving embedding methods, such as modularized nonnegative matrix factorization (M-NMF) \cite{wang2017community} and evolutionary algorithm (EA)-based approaches \cite{li2020evolutionary,Wang2020Surrogate}, have demonstrated the value of integrating community information. However, M-NMF suffers from high computational complexity due to iterative matrix decomposition and lacks support for incremental learning, making it unsuitable for large-scale or dynamic networks. Similarly, EA-based methods, such as EA-NECommunity \cite{li2020evolutionary}, while capable of preserving both local proximity and community structures, are computationally intensive. These limitations highlight the need for scalable, dynamic, and community-aware embedding methods that strike a balance between computational efficiency and representational richness.

\begin{figure}[h!]
    \centering
    \includegraphics[width=0.5\textwidth]{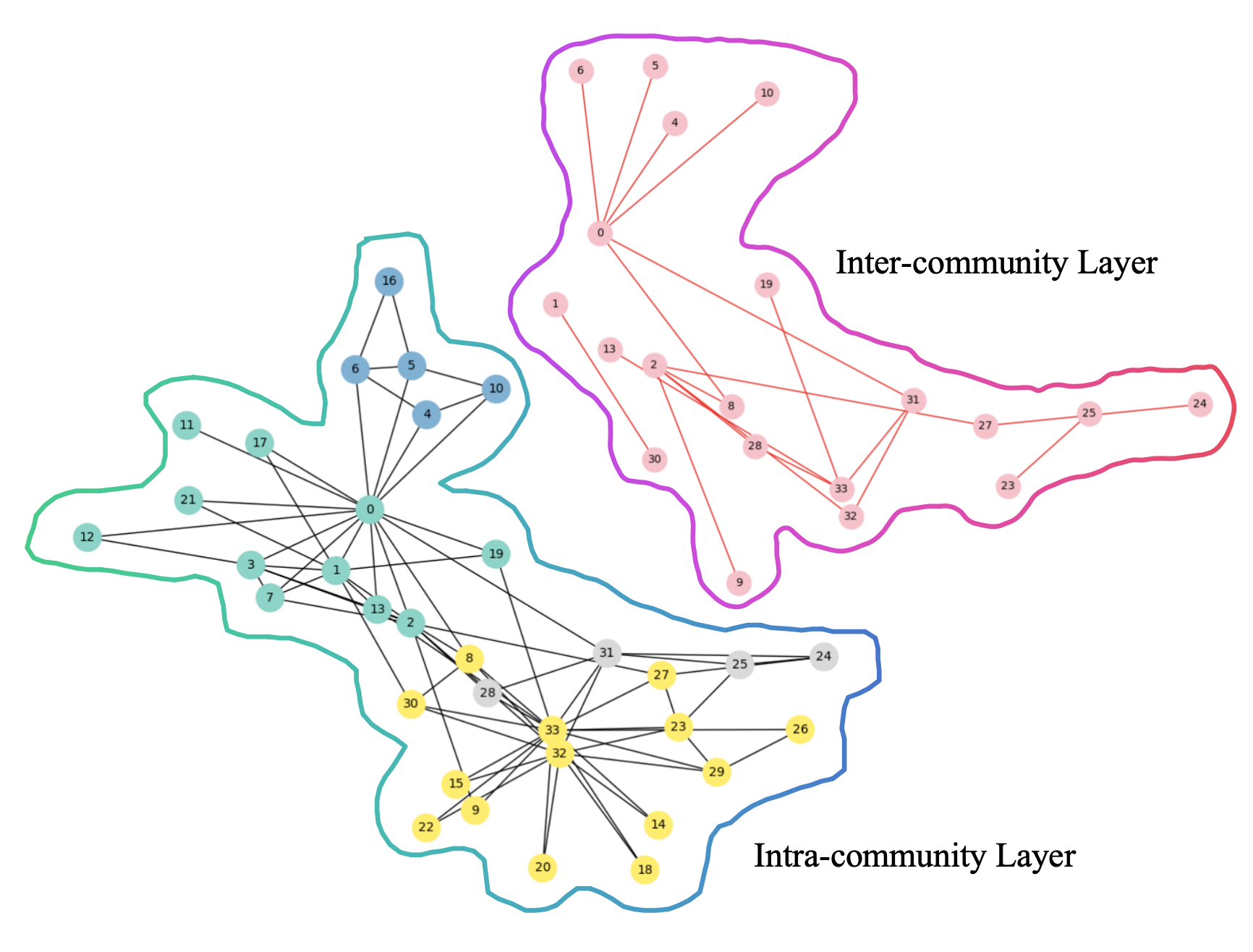} % Replace with actual path
    \caption{Illustration of the hierarchical organization in the Zachary’s Karate Club network. The intra-community level (colored groups) captures dense connections within communities, while the inter-community level (pink edges) highlights sparser connections linking different communities. This structure forms the basis for TLWalk's two-layer embedding approach.}
    \label{fig:karate_club_structure}
\end{figure}

The random walk mechanism used in DeepWalk and node2vec is simple yet powerful for extracting local structural and positional information in graphs. Their inherent flexibility allows for incremental updates and scalability to large graphs. Traditional methods often mix local and global walks, resulting in a loss of critical mesoscopic information and an inability to fully leverage community structures. In contrast, we propose Two Layer Walk (TLWalk), a novel graph embedding algorithm designed to overcome the limitations of existing approaches in capturing community structures. TLWalk introduces a hierarchical two-layer design that separates the network into an intra-community layer and an inter-community layer. By independently conducting walks within each layer, TLWalk effectively extracts dense local structures from communities while capturing sparse global relationships across communities. This design allows TLWalk to preserve the nuanced interplay between local and global structures while ensuring communication between layers through “bridging” nodes—key connectors that link communities. This separation and targeted exploration enable TLWalk to achieve a more precise and comprehensive embedding of network structures, addressing gaps left by existing methods. Figure \ref{fig:karate_club_structure} illustrates this hierarchical organization using the Zachary’s Karate Club network \cite{zachary1977karate}. 

This study rigorously evaluates TLWalk through experiments on benchmark datasets spanning social, biological, and ecological domains. The results consistently demonstrate TLWalk’s superior performance in tasks such as link prediction, node classification, and clustering, outperforming state-of-the-art methods while maintaining computational efficiency. Our key contributions are as follows:

\begin{enumerate}
    \item  We propose a graph embedding algorithm that explicitly incorporates community structures into random walk-based methods without using any parameters, effectively bridging local and global perspectives in network analysis.
    \item We theoretically prove the proposed TKWalk can overcome locality bias, thus, can fully extract both local and global topologies. We also theoretically prove the relationship between TLWalk and matrix factorization. 
    \item We validate the versatility of TLWalk, demonstrating significant performance improvements across key tasks, including link prediction, node classification, and clustering.
\end{enumerate}

By addressing critical challenges in graph embedding, this study not only enhances predictive performance but also advances the theoretical and practical development of community-aware embedding algorithms. Leveraging random walk-based methods to effectively extract community structures, TLWalk strikes a balance between computational efficiency and representational richness, establishing itself as a powerful tool for diverse real-world network analysis tasks.

\section*{Methods}

\subsection*{Preliminaries}
A graph is represented as $G = (V, E)$, where $V$ is the set of nodes, and $E$ is the set of edges connecting pairs of nodes. For an undirected graph, $(u, v) \in E$ implies $(v, u) \in E$. The adjacency matrix of $G$ is denoted by $A \in \{0, 1\}^{|V| \times |V|}$, where $A[u, v] = 1$ if $(u, v) \in E$, and $A[u, v] = 0$ otherwise.

The degree of a node $v$, denoted by $d(v)$, is the number of edges incident to $v$:
\begin{equation}
d(v) = \sum_{u \in V} A[v, u].
\end{equation}

A community in a graph refers to a subset of nodes that are densely connected internally but sparsely connected to nodes outside the subset. A community partition of $G$ is denoted by $\mathcal{C} = \{C_1, C_2, \ldots, C_k\}$, where $C_i \subseteq V$, $i = 1,2,...,k$ and $\bigcup_{i=1}^k C_i = V$. 

The quality of a community partition can be evaluated using modularity $Q$ \cite{newman2004finding},  which is defined as:
\begin{equation}
Q = \frac{1}{2|E|} \sum_{u, v \in V} \left[ A[u, v] - \frac{d(u) d(v)}{2|E|} \right] \delta(C_u, C_v),
\end{equation}
where $\delta(C_u, C_v)$ is $1$ if nodes $u$ and $v$ belong to the same community and $0$ otherwise.

Word2Vec \cite{mikolov2013efficient} is a widely used model for learning vector representations of words based on their contextual relationships. In the context of graph embedding, nodes are treated as "words", and random walks over the graph generate "sentences". The Skip-Gram model \cite{mikolov2013efficient}, a common variant, seeks to maximize the likelihood of observing neighboring nodes within context $C$:
\begin{equation}
\mathcal{L} = - \sum_{(u, v) \in C} \log P(v | u),
\end{equation}
where $P(v | u)$ is the probability of observing node $v$ given node $u$ in the random walk sequences.

To assist readers in understanding the symbols used throughout the paper, we provide a summary of notation in Table~\ref{tab:notation}.

\begin{table}[h!]
\centering
\caption{Summary of Notation}
\label{tab:notation}
\resizebox{0.48\textwidth}{!}{%
\begin{tabular}{|c|l|}
\hline
\textbf{Symbol} & \textbf{Description} \\ \hline
$G = (V, E)$ & Graph with node set $V$ and edge set $E$ \\ \hline
$A$ & Adjacency matrix of $G$ \\ \hline
$d(v)$ & Degree of node $v$ \\ \hline
$\mathcal{C}=\{C_1, C_2, ..., C_k\}$ & Community partition of $G$ \\ \hline
$C_i$ & A single community in $\mathcal{C}$ \\ \hline
$G_i = (V_i, E_i)$ & Subgraph induced by nodes in community $C_i$ \\ \hline
$G_c = (V_c, E_c)$ & Inter-community graph with $V_c$ as inter-community nodes \\ \hline
$V_c$ & Set of nodes participating in inter-community connections \\ \hline
$V_i$ & Set of nodes within community $C_i$ \\ \hline
$Q$ & Modularity score for community partition \\ \hline
$L_w$ & Length of a random walk \\ \hline
$R$ & Number of random walks per node \\ \hline
$P_{\text{intra}}(v | u)$ & Transition probability within community $C_i$ \\ \hline
$P_{\text{inter}}(v | u)$ & Transition probability within inter-community graph $G_c$ \\ \hline
$d$ & Dimensionality of the embedding space \\ \hline
\end{tabular}
}
\end{table}

\subsection*{Two Layer Walk Algorithm}
TLWalk is a graph embedding algorithm designed to capture mesoscopic community structures while maintaining scalability and efficiency. By explicitly modeling both intra-community and inter-community relationships, the method ensures robust representations that encode both local and global graph patterns. The algorithm consists of three main components: community detection, hierarchical random walks, and embedding generation using Word2Vec.

\subsubsection*{Community Detection and Modeling}
To identify mesoscopic structures in the graph, we employ the Louvain algorithm for community detection \cite{blondel2008fast}, which optimizes modularity $Q$ to partition nodes into communities. Each community is defined as a densely connected group of nodes, while inter-community edges represent connections between different groups. This leads to two type of key subgraphs:

\textbf{Intra-Community Subgraphs:} For each community $C_i$, we construct a subgraph $G_i = (V_i, E_i)$, where $V_i$ is the set of nodes within community $C_i$. $E_i$ is the set of edges between nodes in $C_i$. These subgraphs capture dense local patterns, ensuring that nodes within the same community are closely embedded.

\textbf{Inter-Community Subgraph:} The inter-community subgraph $G_c = (V_c, E_c)$ captures connections between communities, where $V_c$ is the set of bridging nodes that participate in at least one inter-community edge, which connecting two nodes in different communities. $E_c$ is the set of edges between nodes in $V_c$ that connect different communities. This representation captures the sparse global structure of the network, focusing on the relationships between different communities.

\subsubsection*{Path-Level Random Walks}

Random walks in TLWalk are designed to operate fully in the \textbf{intra-community layer} or the \textbf{inter-community layer} , where the intra-community layer corresponds to the intra-community subgraphs and the inter-community layer corresponds to the inter-community subgraphs. In this way, the path-level complete random walk in each layer ensures a comprehensive exploration of both dense local structures and sparse global relationships. Each walk with length $L_w$ , $w=(w_0, w_1, ..., w_{L_w-1})$, is conducted entirely within a selected layer according to the type of starting node $w_0$. According to the current community division, all nodes are divided into two types:

\textbf{1. Bridging nodes} (\(v \in V_c\)):  For bridging nodes, the walk explores the inter-community layer \(G_c\), capturing sparse global structures across communities. When $t=0$, the walk starts from $v$; that is, $w_0=v$. Then, $w_{t+1}, t=1, 2, ..., L_{w-1}$ is selected with the following probability:    
\begin{equation}
    P_{\text{inter}}(w_{t+1} | w_t) = \frac{A[w_t, w_{t+1}]}{\sum_{v' \in V_c} A[w_t, v']}, \quad w_t, w_{t+1} \in V_c.
\end{equation}

\textbf{2. Non-bridging nodes (Community-exclusive nodes)} (\(v \in (V-V_c)\cap V_{i}\)): For nodes in community $C_i$ and solely connecting to nodes in $C_i$, their walks are primarily restricted to the intra-community layer, ensuring that dense local structures are captured. When \(t=0\), the walk starts from \(v\); that is, \(w_0 = v\). Then, \(w_{t+1}\), $t=1, 2, ..., L_{w-1}$ is selected with the following probability:
\begin{equation}
    P_{\text{intra}}(w_{t+1} | w_t) = \frac{A[w_t, w_{t+1}]}{\sum_{v' \in V_i} A[w_t, v']}, \quad w_t, w_{t+1} \in V_i.
\end{equation}
The next step \(w_{t+1}\) may include bridging nodes in \(V_i\), as they are part of community \(C_i\). This design ensures that while the walk is restricted to community \(C_i\), it captures both dense intra-community structures and the influence of bridging nodes within the community.

 The random walk mechanism, detailed in Algorithm~\ref{alg:hierarchical_random_walks}, seamlessly integrates this layer-specific exploration strategy, allowing the algorithm to adaptively encode the complex interplay between dense local patterns and sparse global relationships.

\begin{algorithm}[H]
\caption{Two Layer Random Walks}
\label{alg:hierarchical_random_walks}
\textbf{Input:} 
Intra-community subgraphs $\{G_1, G_2, \ldots, G_k\}$,\\
Inter-community graph $G_c = (V_c, E_c)$,\\
Walk length $L_w$, Number of walks per node $R$  \\
\textbf{Output:} Corpus of random walks $\mathcal{W}$ for embedding generation

\begin{algorithmic}[1]
\State $\mathcal{W} \gets \emptyset$ \Comment{\textit{Initialize the corpus of walks}}
\For{each node $v \in V$}
    \For{$r \gets 1$ to $R$}
        
        \If{$v \in V_c$ }

            \State $G_{\text{current}} \gets G_c$ 

        \Else
            \State $G_{\text{current}} \gets G_i$, that $v \in C_i$ 
        \EndIf
        \State $w_0 \gets v$ 
        \For{$t \gets 0$ to $L_w - 1$}
            \State Select $w_{t+1}$ randomly from neighbors of $w_t$ in $G_{\text{current}}$
        \EndFor
        \State $\mathcal{W} \gets \mathcal{W} \cup \{w\}$ 
    \EndFor
\EndFor
\State \textbf{return} $\mathcal{W}$
\end{algorithmic}
\end{algorithm}

As can be seen, the number of nodes participating the intra-community layer walk and the inter-community layer walk are determined automatically based on the intrinsic properties of the graph under consideration, without using any parameter to manipulate. \emph{This adjustment reflects the network's structural properties without requiring manual parameters and enables TLWalk to automatically and systematically balance local and global exploration by leveraging both the intra-community and inter-community layers}.

The random walks $\mathcal{W}$ are treated as input sequences for the Word2Vec model, where each sequence corresponds to a sentence and each node to a word. Using the Skip-Gram architecture, Word2Vec maximizes the co-occurrence probability of nodes within a predefined window, learning a low-dimensional vector representation for each node.

\subsection*{Computational Complexity}
The computational efficiency of TLWalk is achieved through modularity,  hierarchical design, making it suitable for large-scale graphs. Below, we analyze the complexity of its components:

The community detection step leverages the Louvain algorithm, which is widely regarded for its efficiency and scalability. The complexity of this algorithm is:
\begin{equation}
O(|E| \log |V|),
\end{equation}
where \(|E|\) is the number of edges and \(|V|\) is the number of nodes in the graph. This logarithmic dependence on \(|V|\) arises from the iterative optimization of modularity. 

Random walks in TLWalk are performed separately within intra-community subgraphs and the inter-community graph, the cost of generating a walk is proportional to the walk length \(L_w\) and the number of walks per node \(R\). The total complexity of this step is:
\begin{equation}
O(R \cdot L_w \cdot |V|).
\end{equation}

Combining the above steps, the overall complexity of TLWalk is:
\begin{equation}
O(|E| \log |V|) + O(R \cdot L_w \cdot |V|).
\end{equation}

\section*{Results}
In this section, we evaluate the proposed TLWalk (TLW) method against six baseline methods: node2vec (N2V) \cite{grover2016node2vec}, DeepWalk (DW) \cite{perozzi2014deepwalk}, LINE \cite{tang2015line}, Struc2Vec (S2V) \cite{ribeiro2017struc2vec}, GraphSage (GS) \cite{hamilton2017graphsage}, and M-NMF (MN) \cite{wang2017community}. 

The implementations of these methods are sourced from publicly available repositories. Specifically, node2vec, DeepWalk, and M-NMF are implemented using the Karate Club library \cite{karateclub2020} (\url{https://github.com/benedekrozemberczki/karateclub}). LINE is implemented based on the repository provided at \url{https://github.com/tangjianpku/LINE}, while Struc2Vec and GraphSage are sourced from their respective repositories at \url{https://github.com/leoribeiro/struc2vec} and \url{https://github.com/williamleif/GraphSAGE}.

All graph datasets used in our experiments are publicly available at NetworkRepository \cite{nr2015} (\url{https://networkrepository.com}) and the Stanford Large Network Dataset Collection \cite{snap2016} (\url{https://snap.stanford.edu/data}).
\subsection*{Experiment 1: Link Prediction}
Link prediction \cite{liben2007link} assesses the capability of embedding methods to infer missing edges in a graph, which is essential for understanding network dynamics and uncovering hidden relationships. We conduct experiments on six datasets spanning diverse domains, including \textbf{social networks} (soc-hamsterster, fb-pages-food, ego-Facebook), \textbf{biological networks} (bio-DM-LC, bio-WormNet-v3), and \textbf{ecological networks} (aves-weaver-social), as summarized in Table~\ref{tab:dataset_statistics}.

\begin{table}[h!]
\centering
\caption{Dataset Statistics}
\label{tab:dataset_statistics}
\resizebox{0.48\textwidth}{!}{%
\begin{tabular}{lccccc}
\toprule
\textbf{Dataset} & \(|V|\) & \(|E|\) & \(d_{\text{avg}}\) & \(K_{\text{avg}}\) & \(T_{\text{frac}}\) \\ 
\midrule
fb-pages-food      & 620   & 2091   & 6.7      & 0.33   & 0.22 \\ 
ego-Facebook       & 4039  & 88234  & 43.69    & 0.60   & 0.26 \\ 
soc-hamsterster    & 2000  & 16097  & 16.10    & 0.5375 & 0.2314 \\ 
aves-weaver-social & 117   & 304    & 5.20     & 0.6924 & 0.5747 \\ 
bio-DM-LC          & 483   & 997    & 4.13     & 0.1047 & 0.0551 \\ 
bio-WormNet-v3     & 2274  & 78328  & 68.89    & 0.8390 & 0.7211 \\ 
\bottomrule
\end{tabular}%
}
\begin{flushleft}
\footnotesize{\textbf{Notation:} 
\(|V|\): Number of nodes; 
\(|E|\): Number of edges; 
\(d_{\text{avg}}\): Average node degree; 
\(K_{\text{avg}}\): Average clustering coefficient; 
\(T_{\text{frac}}\): Fraction of closed triangles.
}
\end{flushleft}
\end{table}

\subsubsection*{Experimental Setup}
We split each graph into 70\% edges for training and 30\% edges for testing. Node embeddings are learned using the training graph, where edges corresponding to the testing edges are removed to ensure that the testing set remains unseen during embedding learning. For each split (training or testing), positive and negative samples are generated separately based on the corresponding subset of edges. Positive samples consist of edges in the graph, while negative samples are non-existent edges randomly sampled from the graph.

For TLW, DW, and N2V, we set the hyperparameters to \(L_w = 80\) (walk length), \(R = 10\) (number of walks per node), \(d = 128\) (embedding dimension). The performance of all algorithms is evaluated using the logistic regression, and AUC (Area Under Curve).  Edge embeddings are computed by applying the element-wise Hadamard product (\(v_i \odot v_j\)) on the node embeddings of nodes \(v_i\) and \(v_j\). All experiments are repeated for 10 random seeds, and the average AUC scores are reported in Table~\ref{tab:auc_results}. 

\subsubsection*{Results and Analysis}
\begin{figure}[h!]
\centering
\includegraphics[width=0.48\textwidth]{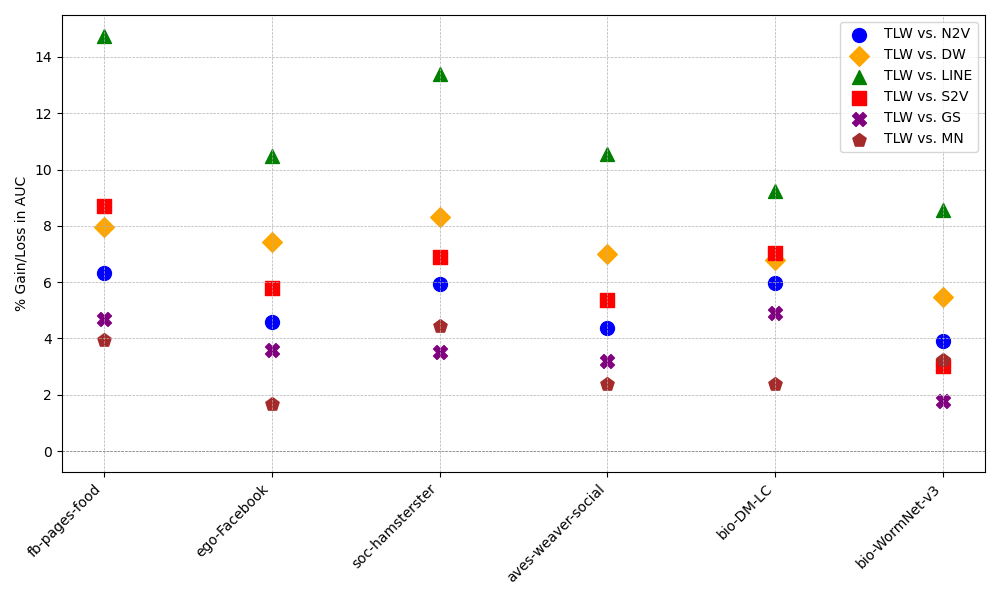}
\caption{AUC Gain/Loss of TLW over Baseline Methods. TLW achieves significant gains over baseline methods, particularly on fb-pages-food and soc-hamsterster, while maintaining consistent performance across diverse datasets.}
\label{fig:Figure1}
\end{figure}

Table~\ref{tab:auc_results} presents the AUC scores for TLW and baseline methods. TLW consistently achieves the highest scores across all datasets, demonstrating its effectiveness in preserving both local and global graph features. On \textit{bio-WormNet-v3}, TLW achieves an AUC of 0.9915, surpassing M-NMF by 3.2\%, while on \textit{ego-Facebook}, TLW achieves 0.9885, showcasing robust performance on large-scale social networks. Notably, TLW excels on datasets with strong community structures, such as \textit{soc-hamsterster} (AUC 0.9122 vs. 0.861 for the best baseline). Illustrated in Figure~\ref{fig:Figure1}, TLW achieves significant gains over baseline methods, particularly on \textit{fb-pages-food} and \textit{soc-hamsterster}, while maintaining consistent performance across diverse datasets.

\begin{table}[h!]
\centering
\caption{Average AUC Scores for Link Prediction (bold numbers represent the best results).}
\label{tab:auc_results}
\resizebox{0.48\textwidth}{!}{%
\begin{tabular}{lccccccc}
\toprule
\textbf{Dataset} & \textbf{TLW} & \textbf{N2V} & \textbf{DW} & \textbf{LINE} & \textbf{S2V} & \textbf{GS} & \textbf{MN} \\ 
\midrule
fb-pages-food      & \textbf{0.9319} & 0.8764 & 0.8632 & 0.8121 & 0.8573 & 0.8902 & 0.8964 \\ 
ego-Facebook       & \textbf{0.9885} & 0.9452 & 0.9203 & 0.8947 & 0.9345 & 0.9541 & 0.9721 \\ 
soc-hamsterster    & \textbf{0.9122} & 0.8610 & 0.8423 & 0.8045 & 0.8534 & 0.8812 & 0.8734 \\ 
aves-weaver-social & \textbf{0.9114} & 0.8732 & 0.8517 & 0.8245 & 0.8651 & 0.8832 & 0.8903 \\ 
bio-DM-LC          & \textbf{0.9337} & 0.8810 & 0.8743 & 0.8547 & 0.8723 & 0.8901 & 0.9121 \\ 
bio-WormNet-v3     & \textbf{0.9915} & 0.9543 & 0.9402 & 0.9134 & 0.9624 & 0.9742 & 0.9603 \\ 
\bottomrule
\end{tabular}
}
\end{table}

These results highlight TLW’s ability to capture both mesoscopic community structures and global connectivity patterns, achieving superior performance and demonstrating its robustness and adaptability in link prediction tasks across diverse network topologies.

\subsection*{Experiment 2: Node Clustering and Classification}

We evaluated the performance of TLW on two fundamental tasks: \textit{Node Clustering} and \textit{Node Classification}, using nine diverse real-world networks that include the label of community. These experiments demonstrate the effectiveness of TLW compared to six baseline methods across various domains.

\subsubsection*{Datasets and Experimental Setup}

The datasets span academic, social, and political domains. \textbf{WebKB}: Academic networks from Cornell, Texas, Washington, and Wisconsin~\cite{craven1998learning, webkbdata}, comprising 877 webpages and 1608 edges, divided into five ground-truth communities per university. \textbf{Polblogs}: A political blog network with 1222 nodes and 16,715 edges, categorized into two communities (liberal and conservative)~\cite{adamic2005political}. \textbf{Facebook Social Networks}: University social networks, including Amherst, Hamilton, Mich, and Rochester, where ground-truth communities correspond to users' class years~\cite{traud2012social}.

Embeddings were generated using TLW and baseline methods with a uniform embedding dimension (\(d = 128\)). For \textit{Node Clustering}, K-means clustering~\cite{lloyd1982least} was applied to the learned representations, and clustering accuracy was calculated. For \textit{Node Classification}, a logistic regression model~\cite{bishop2006pattern} classified nodes into labels, with 80\% of nodes used for training and 20\% for testing. Each task was repeated five times, and average results are reported. This evaluation approach is consistent with prior studies on graph embeddings.

\subsubsection*{Node Clustering Results}

Table~\ref{tab:clustering_accuracy} summarizes the clustering accuracy across datasets. TLW consistently achieves the highest accuracy on most datasets. Notably, TLW achieves 85.17\% accuracy on the \textit{Polblogs} dataset, surpassing the next best method (N2V) by 0.4\%. On academic datasets, such as \textit{Cornell} and \textit{Texas}, TLW outperforms the second-best baseline by 4.1\% and 1.9\%, respectively. On smaller datasets, such as \textit{Rochester}, TLW demonstrates a notable lead, achieving 39.35\% accuracy compared to 33.80\% by GS. These results highlight TLW's strength in preserving community structures across diverse graph topologies.

\begin{table}[h!]
\centering
\caption{Accuracy (\%) of Node Clustering (bold numbers represent the best results).}
\label{tab:clustering_accuracy}
\resizebox{0.48\textwidth}{!}{%
\begin{tabular}{lccccccc}
\toprule
\textbf{Methods} & \textbf{TLW} & \textbf{N2V} & \textbf{DW} & \textbf{LINE} & \textbf{S2V} & \textbf{GS} & \textbf{MN} \\ 
\midrule
Cornell     & \textbf{44.70} & 34.36 & 32.82 & 42.56 & 31.38 & 38.85 & 43.05 \\ 
Texas       & \textbf{65.06} & 50.27 & 37.97 & 55.61 & 40.64 & 45.29 & 63.10 \\ 
Washington  & \textbf{60.65} & 41.74 & 35.65 & 53.48 & 38.70 & 46.52 & 59.57 \\ 
Wisconsin   & \textbf{48.64} & 35.47 & 34.34 & 43.77 & 35.09 & 41.60 & 45.66 \\ 
Polblogs    & \textbf{85.17} & 84.83 & 52.68 & 63.88 & 57.38 & 73.42 & 82.82 \\ 
Amherst     & \textbf{48.54} & 41.66 & 10.34 & 44.38 & 23.36 & 46.41 & 47.25 \\ 
Hamilton    & \textbf{43.14} & 35.41 & 10.15 & 31.30 & 33.47 & 38.81 & 42.49 \\ 
Mich        & 32.61 & 14.05 & 11.66 & 14.63 & 15.58 & \textbf{35.12} & 31.50 \\ 
Rochester   & \textbf{39.35} & 18.00 & 7.94  & 16.86 & 12.88 & 33.80 & 38.09 \\ 
\bottomrule
\end{tabular}
}
\end{table}

\subsubsection*{Node Classification Results}

Table~\ref{tab:classification_accuracy} presents the classification accuracy. TLW outperforms all baseline methods in nine datasets. On the \textit{Amherst} dataset, TLW achieves 93.74\%, surpassing the next best method (GS) by 2.2\%. Similarly, on \textit{Polblogs}, TLW achieves 91.20\%, showcasing its ability to leverage both local and global features. On challenging datasets, such as \textit{Texas}, TLW maintains a competitive edge, achieving 74.89\% accuracy with a 1.7\% lead over MN. These results demonstrate that TLW produces high-quality embeddings that excel in classification tasks, effectively capturing both mesoscopic and global graph structures.

\begin{table}[h!]
\centering
\caption{Accuracy (\%) of Node Classification (bold numbers represent the best results).}
\label{tab:classification_accuracy}
\resizebox{0.48\textwidth}{!}{%
\begin{tabular}{lccccccc}
\toprule
\textbf{Methods} & \textbf{TLW} & \textbf{N2V} & \textbf{DW} & \textbf{LINE} & \textbf{S2V} & \textbf{GS} & \textbf{MN} \\ 
\midrule
Cornell     & \textbf{49.69} & 38.46 & 24.10 & 44.62 & 27.69 & 45.38 & 47.18 \\ 
Texas       & \textbf{74.89} & 51.05 & 22.63 & 73.16 & 34.21 & 68.42 & 70.00 \\ 
Washington  & \textbf{64.67} & 53.78 & 24.44 & 50.22 & 25.33 & 52.00 & 63.56 \\ 
Wisconsin   & \textbf{64.62} & 44.62 & 26.15 & 51.54 & 28.46 & 59.62 & 61.15 \\ 
Polblogs    & \textbf{91.20} & 84.03 & 64.77 & 80.87 & 70.02 & 89.60 & 90.67 \\ 
Amherst     & \textbf{93.74} & 89.73 & 41.59 & 87.99 & 72.51 & 91.46 & 92.00 \\ 
Hamilton    & \textbf{93.36} & 87.27 & 39.95 & 87.27 & 78.64 & 91.64 & 92.92 \\ 
Mich        & \textbf{64.15} & 61.98 & 25.44 & 60.75 & 50.09 & 60.79 & 62.26 \\ 
Rochester   & \textbf{89.28} & 83.65 & 34.78 & 84.23 & 46.04 & 85.47 & 87.18 \\ 
\bottomrule
\end{tabular}
}
\end{table}

\subsection*{Experiment 3: Community Detection on LFR Benchmark Networks}
To evaluate the performance of node embedding in detecting community structures \cite{kojaku2024network,ghasemian2021community}, we conducted experiments on synthetic networks generated using the \textbf{LFR benchmark} model~\cite{lancichinetti2008benchmark}. The LFR model allows for realistic community structures with heterogeneous degree and community size distributions, which are essential for evaluating the robustness and scalability of graph embedding methods in practical scenarios.

\subsubsection*{Experimental Setup}

Synthetic networks are generated using the LFR benchmark model as follows: Each network consists of \( |V| = 10^4 \) nodes with an average degree \( \langle k \rangle = 10 \), ensuring a moderate connectivity level. The degree distribution follows a power law with degree exponents  $\tau_1$ = 2.1  and  $\tau_1$ = 3.0 , where  $\tau_1$ = 2.1  represents a highly heterogeneous distribution with a few nodes having significantly higher degrees than most others, while  $\tau_1$ = 3.0  corresponds to a less heterogeneous distribution with a more balanced degree spread across nodes. Nodes are divided into multiple communities with a predefined size distribution. To control the ratio of inter-community edges relative to the total degree of a node, we use the \textit{mixing parameter} \( \mu \), defined as:
\begin{equation}
\mu = \frac{|V|\cdot p_{\text{out}}}{\langle k \rangle},
\end{equation}
where \( p_{\text{out}} \) is the probability of connecting to nodes outside the community. The value of \( \mu \) ranges from 0.1 to 1.0: \( \mu = 0.1 \) corresponds to well-separated communities with minimal inter-community edges, while \( \mu = 1.0 \) represents fully blended communities with maximal inter-community edges.

\paragraph{Clustering and Matching}
The embeddings were used for \textit{node clustering} using the {K-means algorithm~\cite{lloyd1982least}, where the number of clusters \( K \) was set to the true number of communities in the LFR network. To evaluate the quality of the detected communities, we matched the detected clusters with the ground-truth communities using the Hungarian matching algorithm~\cite{kuhn1955hungarian}, which ensures optimal alignment between the predicted and true labels.

\paragraph{Performance Metric}
We measure the performance of varying algorithms using the {element-centric similarity \( S \)~\cite{zhang2016community}, a widely used metric for evaluating community detection. The similarity \( S \) quantifies how well the detected communities align with the ground-truth partition: \( S > 0 \) indicates detectable communities, while \( S = 0 \) is the baseline for random assignments.

\subsubsection*{Results and Analysis}

\textbf{Figure~\ref{fig:lfr_experiment}} presents the performance comparison across all methods. 
\begin{figure*}[htbp]
    \centering
    \includegraphics[width=0.8\textwidth]{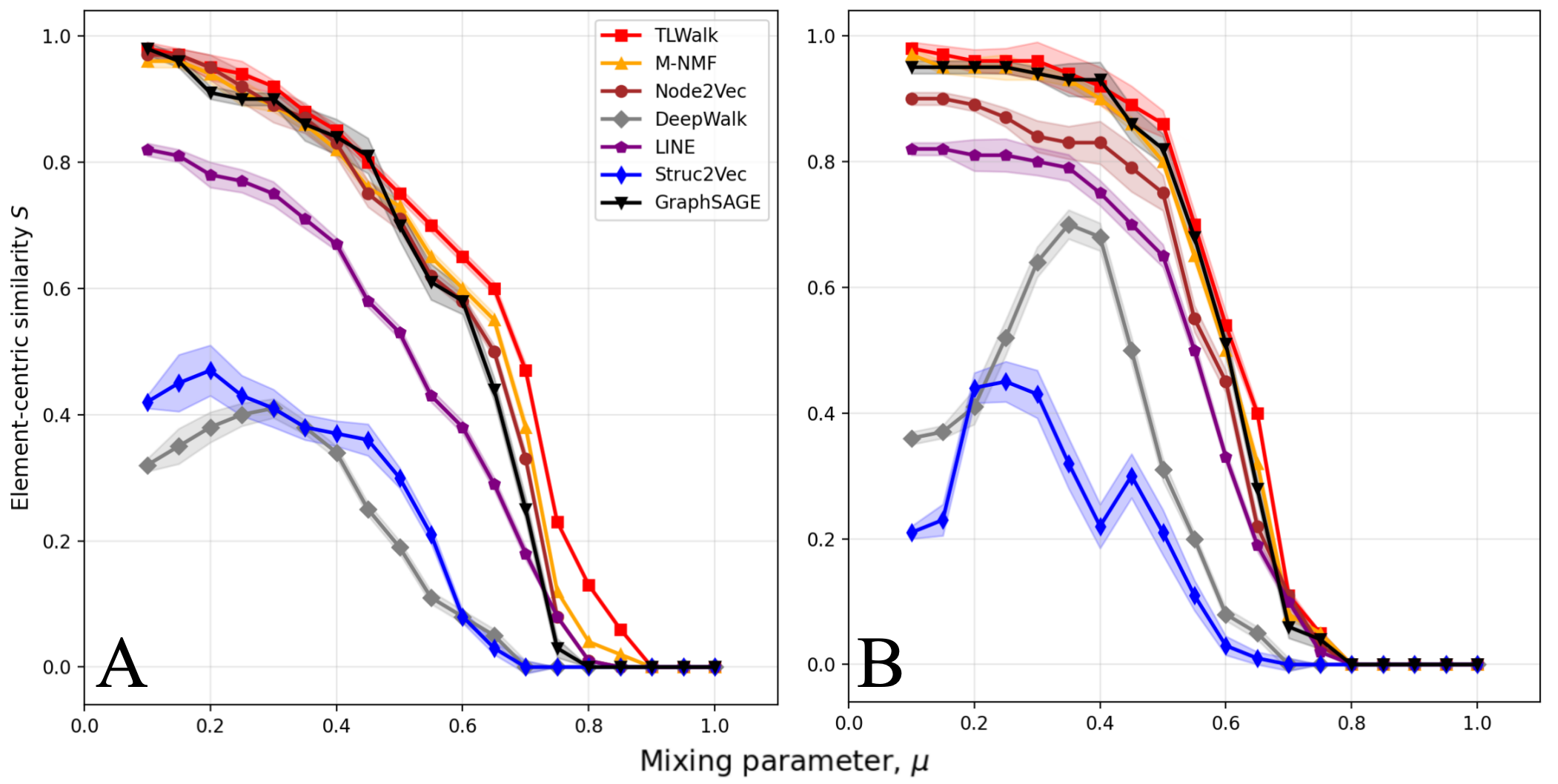}
    \caption{Performance comparison of TLWalk and baseline methods on LFR benchmark networks as a function of the mixing parameter \( \mu \). 
    (A) \( \tau_1 = 2.1 \) (highly heterogeneous degree distribution), 
    (B) \( \tau_1 = 3.0 \) (less heterogeneous degree distribution). TLWalk consistently achieves superior performance across all levels of \( \mu \).}
    \label{fig:lfr_experiment}
\end{figure*}

\begin{itemize}
    \item \textbf{Case A (\( \tau_1 = 2.1 \))}: As shown in Figure~\ref{fig:lfr_experiment}A, TLWalk consistently outperforms all baseline methods across the entire range of the mixing parameter \( \mu \). At low values of \( \mu \), where communities are well-separated, TLWalk achieves near-perfect similarity scores \( S \). As \( \mu \) increases, indicating stronger inter-community blending, TLWalk maintains its superior performance, while baseline methods such as Struc2Vec and DeepWalk degrade significantly earlier. This demonstrates TLWalk's robustness in networks with highly heterogeneous degree distributions.

    \item \textbf{Case B (\( \tau_1 = 3.0 \))}: In Figure~\ref{fig:lfr_experiment}B, where the degree distribution is less heterogeneous, all methods exhibit improved performance for \( \mu < 0.65 \) compared to \( \tau_1 = 2.1 \). However, TLWalk continues to maintain a distinct advantage, particularly in the mid-range of \( \mu \) (0.3--0.6), where community detectability becomes increasingly difficult. This highlights TLWalk's ability to adapt and perform consistently across varying network conditions.
\end{itemize}

These results demonstrate TLWalk's robustness and effectiveness in community detection tasks, particularly in challenging network conditions with varying degree distributions and mixing levels.

\section*{Theoretical Analysis}

In this section, we provide a theoretical foundation for the proposed TLWalk method by addressing two critical aspects of its design. First, we analyze how TLWalk mitigates the locality bias inherent in traditional random walks. By separating intra- and inter-community transitions, TLWalk ensures balanced exploration of dense local structures and sparse global relationships, enabling embeddings to capture both local and global network properties better. Second, we interpret TLWalk from the perspective of matrix factorization. We demonstrate that the two-layer random walk process in TLWalk corresponds to a novel matrix factorization form, which integrates both intra- and inter-community dynamics through a hierarchical transition mechanism. These analyses highlight TLWalk’s theoretical advantages in encoding community-aware structures, providing insights into its effectiveness in network representation learning.

\subsection*{Analysis 1: Overcoming Locality Bias in Traditional Random Walks}

Let a sequence of nodes generated by a traditional random walk be \( \{w_0, w_1, ..., w_{L_w-1}\} \), where \( w_t \in V \) and \( t = 0, 1, ..., L_{w-1} \). At time \( t+1 \), \( t = 0, 1, ..., L_{w-2} \), the walk moves to a neighbor of \( w_t \) with probability \( 1/d(w_t) \). Denote \( P_{uv} \) as the probability of moving from node \( u \) to node \( v \) in one step, which is defined as:
\[
P_{uv}= 
\begin{cases}
1/d(u),& \text{if } (u,v) \in E, \\
0,              & \text{otherwise.}
\end{cases}
\]

Suppose \( w_0 = u \) and \( w_t = v \), where \( u, v \in V \) and \( w_0, w_1, ..., w_{t-1} \neq v \). The probability that the walk starts at \( u \) and visits \( v \) for the first time at time \( t \) is given by:
\[
r^t_{uv} = \sum_{j \in \mathcal{N}(u)} P_{uj} \cdot r^{t-1}_{jv} = \frac{1}{d(u)} \sum_{j \in \mathcal{N}(u)} r^{t-1}_{jv},
\]
where \( \mathcal{N}(u) \) is the set of neighbors of node \( u \). This equation shows that \( r^t_{uv} \) is the mean of the \((t-1)\)-step probabilities \( r^{t-1}_{jv} \) over all neighbors of \( u \). The reliance on all neighbors dilutes the transition probability towards distant nodes, reinforcing locality bias.

\subsubsection*{Enhanced Global Accessibility in TLWalk}

\textbf{Lemma 1.} In TLWalk, let nodes \( u \) and \( v \) belong to two different communities, and \( u \), \(v \in V_c \) are  bridging nodes. The probability that the walk starts from \( u \) and visits \( v \) for the first time at time \( t \) satisfies:
\begin{equation}
r^t_{uv} \geq \frac{1}{d(u)} \sum_{j \in \mathcal{N}(u)} r^{t-1}_{jv},
\end{equation}
with equality when all neighbors of \( u \) are bridging nodes (\( \mathcal{N}(u) = \mathcal{N}_{\text{inter}}(u) \)).

\textbf{Proof.} Partition \( \mathcal{N}(u) \) into two disjoint subsets:
\begin{itemize}
    \item \( \mathcal{N}_{\text{inter}}(u) = \mathcal{N}(u) \cap V_c \), the set of inter-community neighbors (bridging nodes),
    \item \( \mathcal{N}_{\text{intra}}(u) = \mathcal{N}(u) - \mathcal{N}_{\text{inter}}(u) \), the set of intra-community neighbors.
\end{itemize}

The total probability of transitioning to neighbors of \( u \) can be expressed as:
\begin{equation}
\sum_{j \in \mathcal{N}(u)} r^{t-1}_{jv} = \sum_{j \in \mathcal{N}_{\text{inter}}(u)} r^{t-1}_{jv} + \sum_{j \in \mathcal{N}_{\text{intra}}(u)} r^{t-1}_{jv}.
\end{equation}

In TLWalk, intra-community transitions cannot reach nodes in different communities. Thus, for any \( j \in \mathcal{N}_{\text{intra}}(u) \), \( r^{t-1}_{jv} = 0 \). The total probability simplifies to:
\begin{equation}
\sum_{j \in \mathcal{N}(u)} r^{t-1}_{jv} = \sum_{j \in \mathcal{N}_{\text{inter}}(u)} r^{t-1}_{jv}.
\end{equation}

The transition probability \( r^t_{uv} \) from \( u \) to \( v \) in TLWalk is:
\begin{equation}
r^t_{uv} = \frac{1}{|\mathcal{N}_{\text{inter}}(u)|} \sum_{j \in \mathcal{N}_{\text{inter}}(u)} r^{t-1}_{jv}.
\end{equation}

Since \( |\mathcal{N}_{\text{inter}}(u)| \leq d(u) \), it follows that:
\begin{equation}
\frac{1}{|\mathcal{N}_{\text{inter}}(u)|} \geq \frac{1}{d(u)}.
\end{equation}

Substituting this into the equation for \( r^t_{uv} \), we obtain:
\begin{equation}
r^t_{uv} = \frac{1}{|\mathcal{N}_{\text{inter}}(u)|} \sum_{j \in \mathcal{N}_{\text{inter}}(u)} r^{t-1}_{jv} \geq \frac{1}{d(u)} \sum_{j \in \mathcal{N}(u)} r^{t-1}_{jv}.
\end{equation}

Equality holds when all neighbors of \( u \) are bridging nodes (\( \mathcal{N}(u) = \mathcal{N}_{\text{inter}}(u) \)).

\textbf{Lemma 1} demonstrates that TLWalk reduces locality bias by prioritizing inter-community transitions for bridging nodes. Unlike traditional random walks, which distribute transition probabilities evenly among all neighbors, TLWalk separates intra- and inter-community contributions, enabling effective exploration of global structures. This mechanism ensures embeddings that capture both local and global network properties, making TLWalk well-suited for tasks requiring awareness of inter-community relationships.

\subsection*{Analysis 2: Matrix Factorization Perspective of TLWalk}

The Skip-Gram with negative sampling (SGNS) framework in network embedding methods such as DeepWalk \cite{perozzi2014deepwalk} and node2vec \cite{grover2016node2vec} has been shown to implicitly factorize a pointwise mutual information (PMI) matrix \cite{levy2014neural,Qiu:2018}. TLWalk extends this perspective by incorporating a hierarchical, two-layer random walk that explicitly encodes both intra-community and inter-community transitions, resulting in a novel matrix factorization form. 

\subsubsection*{Two-Layer Transition Matrices}

In TLWalk, the probability of transitioning between two nodes during the random walk is governed by two distinct components: \textit{intra-community transitions} (within communities) and \textit{inter-community transitions} (across communities). To model this, we introduce two matrices:  
1) a \textbf{block-diagonal matrix} \( M_I \in \mathbb{R}^{|V| \times |V|} \) that represents transitions within individual communities, and  
2) an \textbf{extended matrix} \( M_C  \in \mathbb{R}^{|V| \times |V|}\) that captures transitions between communities via bridging nodes.

\paragraph{Intra-community transition matrix (\( M_I \)).}  
The matrix \( M_I \) represents the normalized intra-community transitions, structured as a block-diagonal matrix. Each block \( D_i^{-1} A_i \) corresponds to a community \( C_i \), where:
\begin{itemize}
    \item \( A_i \in \mathbb{R}^{|V_i| \times |V_i|} \) is the adjacency matrix of the subgraph \( G_i = (V_i, E_i) \), induced by the set of nodes \( V_i \) and edges \( E_i \) within community \( C_i \).
    \item \( D_i^{-1} \) is the inverse degree matrix of \( A_i \), ensuring row normalization of transition probabilities within each community.
\end{itemize}

The resulting block-diagonal structure of \( M_I \) is expressed as:
\begin{equation}
M_I =
\begin{pmatrix}
D_1^{-1} A_1 & 0 & \cdots & 0 \\
0 & D_2^{-1} A_2 & \cdots & 0 \\
\vdots & \vdots & \ddots & \vdots \\
0 & 0 & \cdots & D_k^{-1} A_k
\end{pmatrix}.
\end{equation}

This formulation confines random walks within individual communities \( C_1, C_2, \dots, C_k \), effectively capturing the dense intra-community relationships while isolating transitions between distinct communities.

\paragraph{Inter-community transition matrix (\( M_C \)).}  
The matrix \( M_C \) captures transitions between inter-community bridging nodes. To construct \( M_C \), we first normalize the adjacency matrix \( A_c \) of the subgraph \( G_c = (V_c, E_c) \), where \( V_c \) is the set of bridging nodes. The normalized matrix is defined as:
\begin{equation}
A_c^{(n)} = D_c^{-1} A_c,
\end{equation}
where \( D_c \) is the degree matrix of \( A_c \), ensuring that the rows of \( A_c^{(n)} \) sum to one. Here, \( A_c^{(n)} \in \mathbb{R}^{|V_c| \times |V_c|} \) represents the normalized transition probabilities restricted to the bridging nodes.

To integrate \( M_C \) with \( M_I \), we align the node order in \( M_C \) to match the node ordering used in \( M_I \). Specifically, the rows and columns of \( M_C \) are arranged such that:
\begin{itemize}
    \item Nodes within the same community are grouped together, maintaining the block-diagonal structure of \( M_I \),
    \item Rows and columns corresponding to non-bridging nodes \( w \notin V_c \) are filled with zeros.
\end{itemize}

This alignment ensures that both \( M_C \) and \( M_I \) are compatible for subsequent matrix operations, such as multi-step transition summation. The resulting \( M_C \) can be visualized as follows:

\begin{equation}
M_C =
\begin{pmatrix}
0 & 0 & 0 & 0 & 0 & \cdots & 0 \\
0 & A_c^{(n)}[v_2, v_2] & 0 & 0 & A_c^{(n)}[v_2, v_5] & \cdots & 0 \\
0 & 0 & 0 & 0 & 0 & \cdots & 0 \\
0 & 0 & 0 & 0 & 0 & \cdots & 0 \\
0 & A_c^{(n)}[v_5, v_2] & 0 & 0 & A_c^{(n)}[v_5, v_5] & \cdots & 0 \\
\vdots & \vdots & \vdots & \vdots & \vdots & \ddots & \vdots \\
0 & 0 & 0 & 0 & 0 & \cdots & 0
\end{pmatrix}.
\end{equation}

\noindent In this representation:
\begin{itemize}
    \item \( A_c^{(n)} \) is the normalized adjacency matrix of the bridging nodes \( V_c \),
    \item Rows and columns are explicitly aligned to the node order in \( M_I \), ensuring compatibility for operations like matrix addition,
    \item Rows and columns corresponding to non-bridging nodes \( w \notin V_c \) are zeroed out to preserve the structure.
\end{itemize}

This design ensures that \( M_C \) and \( M_I \) remain consistent in terms of node alignment. By isolating inter-community transitions and aligning them to the global node order of \( M_I \), \( M_C \) facilitates a seamless integration of intra-community and inter-community dynamics in the TLWalk framework.

\subsubsection*{Matrix Factorization of TLWalk}

\textbf{Lemma 2: }  Let \( M_C \) and \( M_I \) represent the inter-community and intra-community transition matrices, respectively. The embeddings learned by TLWalk using 
Skip-Gram with negative sampling correspond to a low-rank factorization of the following shifted Pointwise Mutual Information (PMI) matrix:
\begin{equation}
\log \left( \text{vol}(G) \cdot \left( \frac{1}{T} \sum_{r=1}^T \left( M_C^r + M_I^r \right) \right) D^{-1} \right) - \log k,
\end{equation}
where:
\begin{itemize}
    \item \text{vol}(G) = $\sum_i \sum_j A_{i,j}$, \text{is the volume of a graph}\, $G$,
    \item \( T \) is the context window size,
    \item \( M_C^r \) and \( M_I^r \) are the \( r \)-step powers of \( M_C \) and \( M_I \), respectively,
    \item \( D^{-1} \) is the inverse degree matrix of $G$,
    \item \( k \) is the negative sampling parameter.
\end{itemize}

\noindent\textbf{Proof.}  
We begin with the Skip-Gram objective, which seeks to maximize the probability of observing a context node \( c \) given a target node \( w \). Under the TLWalk framework, random walks are independently applied to the inter-community transition matrix \( M_C \) and the intra-community transition matrix \( M_I \).

\paragraph{Step 1: Joint Probability.}  
The joint probability \( P(w, c) \) of observing a node \( w \) with context \( c \) is defined by the stationary distribution \( \pi(w) \) and the \( T \)-step transition probabilities:
\begin{equation}
P(w, c) \approx \pi(w) \cdot \frac{1}{T} \sum_{r=1}^T \left( M_C^r + M_I^r \right)[w, c],
\end{equation}
where:
\begin{itemize}
    \item \( \pi(w) \): The stationary probability of node \( w \), given by \( \pi(w) = \frac{d{(w)}}{\text{vol}(G)} \), where \( d{(w)} \) is the degree of \( w \) and \( \text{vol}(G) \) is the graph volume.
    \item \( M_C^r \) and \( M_I^r \): The \( r \)-step transition probability matrices for \( M_C \) (inter-community transitions) and \( M_I \) (intra-community transitions), respectively.
\end{itemize}

\paragraph{Step 2: Marginal Probabilities.}  
The marginal probability \( P(w) \) of a node \( w \) under the stationary distribution is given by:
\begin{equation}
P(w) = \pi(w) = \frac{d{(w)}}{\text{vol}(G)}.
\end{equation}
Similarly, the marginal probability \( P(c) \) of context \( c \) is:
\begin{equation}
P(c) = \pi(c) = \frac{d{(c)}}{\text{vol}(G)}.
\end{equation}

\paragraph{Step 3: Pointwise Mutual Information (PMI).}  
The PMI between a target node \( w \) and a context node \( c \) is defined as:  
\begin{equation}
\label{PMI}
PMI(w, c) = \log \frac{P(w, c)}{P(w)P(c)}.
\end{equation}
Substituting \( P(w, c) \), \( P(w) \), and \( P(c) \) into Eq.(\ref{PMI}), we obtain:
\begin{equation}
PMI(w, c) \approx \log \frac{\pi(w) \cdot \frac{1}{T} \sum_{r=1}^T \left( M_C^r + M_I^r \right)[w, c]}{\pi(w) \pi(c)}.
\end{equation}
Simplifying:
\begin{equation}
PMI(w, c) \approx \log \left( \frac{1}{T} \sum_{r=1}^T \left( M_C^r + M_I^r \right)[w, c] \cdot \frac{\text{vol}(G)}{d{(c)}} \right).
\end{equation}

\paragraph{Step 4: Matrix Representation.}  
We can represent the PMI matrix in the following matrix form:
\begin{equation}
\log \left( \text{vol}(G) \cdot \left( \frac{1}{T} \sum_{r=1}^T \left( M_C^r + M_I^r \right) \right) D^{-1} \right).
\end{equation}

\paragraph{Step 5: SGNS Objective.}  
The Skip-Gram with negative sampling introduces a shift \( -\log k \), where \( k \) is the number of negative samples. Therefore, the final form of the factorization becomes:
\begin{equation}
\log \left( \text{vol}(G) \cdot \left( \frac{1}{T} \sum_{r=1}^T \left( M_C^r + M_I^r \right) D^{-1} \right) \right) - \log k.
\end{equation}

\textbf{Lemma 2} shows the embeddings learned by TLWalk correspond to the low-rank factorization of the combined \( T \)-step transition probabilities derived from \( M_C \) and \( M_I \), effectively integrating inter-community and intra-community relationships and capturing both local and global relationships without additional parameters.

\section*{Discussion}

In this study, we introduced \textbf{Two Layer Walk (TLWalk)}, a novel graph embedding method that explicitly integrates mesoscopic community structures to address critical gaps in existing graph embedding techniques. By leveraging a hierarchical two-layer random walk mechanism, TLWalk systematically balances intra-community and inter-community relationships, providing a more precise and comprehensive embedding of network structures.

The two-layer design of TLWalk overcomes the limitations of traditional random walk-based methods, such as DeepWalk \cite{perozzi2014deepwalk} and node2vec \cite{grover2016node2vec}, which often mix local and global walks, resulting in the loss of mesoscopic-level structural information. By conducting targeted walks separately within intra- and inter-community layers, TLWalk effectively captures dense local patterns while maintaining sparse global relationships. This balanced exploration ensures the embeddings encode both community-specific structures and broader network topologies, addressing the long-standing locality bias inherent in previous approaches.

Our extensive experiments on benchmark datasets spanning social, biological, and ecological networks highlight TLWalk's superiority. Specifically, TLWalk consistently outperformed six state-of-the-art methods, including modularized approaches like M-NMF \cite{wang2017community} and structure-aware embeddings like Struc2Vec \cite{ribeiro2017struc2vec}. For example, in link prediction tasks, TLWalk achieved significant accuracy improvements on datasets with strong community structures, such as bio-WormNet-v3 and ego-Facebook. Furthermore, TLWalk excelled in node clustering and classification experiments, demonstrating its ability to generalize across domains with varying network properties.

The robustness of TLWalk was further validated on synthetic networks generated using the LFR benchmark. Results showed TLWalk's resilience in detecting communities across networks with both highly heterogeneous degree distributions (\(\tau_1 = 2.1\)) and more balanced degree spreads (\(\tau_1 = 3.0\)). This adaptability across varying network conditions underscores its practical utility in diverse real-world applications.

Despite its strong performance, TLWalk relies on predefined community structures for partitioning the network, which may not always align perfectly with real-world networks. Future research could explore adaptive community detection mechanisms or dynamic embedding approaches that evolve with changes in network topology. Additionally, integrating TLWalk with deep learning frameworks, such as Graph Neural Networks \cite{kipf2017semi}, may enable modeling of complex non-linear relationships, further enhancing its representational power.

In conclusion, TLWalk represents a significant advancement in graph embedding techniques by explicitly incorporating mesoscopic community structures. Its scalability, robustness, and interpretability make it a powerful tool for a wide range of network analysis tasks, including link prediction, node classification, and community detection. The insights gained from this study pave the way for practical applications in social network analysis, bioinformatics, and beyond, providing a foundation for future innovations in community-aware embedding algorithms.

\section*{Data availability}
The datasets used in this study are publicly available at the NetworkRepository~\cite{nr2015} 
(\url{https://networkrepository.com}) and the Stanford Large Network Dataset Collection~\cite{snap2016} 
(\url{https://snap.stanford.edu/data}).

\section*{Code availability}
The code and documentation for reproducing all experimental results are available on GitHub. 
See our archived code at ref.~\cite{TLWalkCode} for reproducing our results, and the up-to-date version at 
\url{https://github.com/leonyuhe/TLWalk/} for replications.

\end{document}